\documentclass[aps, prl, reprint, amsmath, amssymb, groupedaddress]{revtex4-1}

\usepackage{graphicx, subfigure, dcolumn}
\usepackage{bm}

\begin{document}

\title{Universal Quantum Computing with Arbitrary Continuous-Variable Encoding}

\author{Hoi-Kwan Lau}\email[Email address: ]{hklau.physics@gmail.com}
\affiliation{Institute of Theoretical Physics, Ulm University, Albert-Einstein-Allee 11, 89069 Ulm, Germany}
\author{Martin B. Plenio}
\affiliation{Institute of Theoretical Physics, Ulm University, Albert-Einstein-Allee 11, 89069 Ulm, Germany}

\date{\today} 

\begin{abstract}
Implementing a qubit quantum computer in continuous-variable systems conventionally requires the engineering of specific interactions according to the encoding basis states.  
In this work, we present a unified formalism to conduct universal quantum computation with a fixed set of operations but arbitrary encoding.
By storing a qubit in the parity of two or four qumodes, all computing processes can be implemented by basis state preparations, continuous-variable exponential-swap operations, and swap-tests.  Our formalism inherits the advantages that the quantum information is decoupled from collective noise, and logical qubits with different encodings can be brought to interact without decoding.  We also propose a possible implementation of the required operations by using interactions that are available in a variety of continuous-variable systems.  Our work separates the `hardware' problem of engineering quantum-computing-universal interactions, from the `software' problem of designing encodings for specific purposes.  The development of quantum computer architecture could hence be simplified.

\end{abstract}


\maketitle

\textit{Introduction---}  In a wide range of quantum computational tasks, the basic quantity of quantum information is a 
two-level system that can be prepared in an arbitrary superposition state
(qubit) \cite{book:NielsenChuang}.  
If the quantum system
 consists of individually addressable energy eigenstates, such as the internal levels in trapped atoms or the polarisation states of electron spins \cite{Ladd:2010kq}, the qubit bases are most trivially represented by two of such states.  
On the other hand, there are also quantum systems, such as optical modes \cite{Braunstein:2005wr}, mechanical oscillators \cite{Poot:2012fh}, quantised motion of trapped ions \cite{Haffner:2008tg}, and spin ensembles \cite{Tordrup:2008kf,Wesenberg:2009es}, that consist of an abundance of evenly-spaced energy levels.  In these systems, usually referred to as continuous-variable (CV) systems, addressing a particular energy eigenstate is usually challenging.  There is thus no trivial CV representation of a qubit.

Nevertheless, the large Hilbert space of each degree of freedom, usually called a quantum mode (qumode), provides the flexibility for designing a qubit encoding for specific purposes.  Each popular encoding, for which the computational basis states could be Fock states, coherent states, Cat states, superpositions of squeezed states, or else \cite{Chuang:1995ud, Chuang:1997ko, Knill:2001vi, Ralph:2003jk, Anonymous:hr, Leghtas:2013ff,Mirrahimi:2014js, Gottesman:2001vk, Menicucci:2014cx, Chuang:1997dx, Ketterer:2015Code, Michael:2016code}, has its own strength in, e.g., efficiency of initialisation, error-tolerance, or measurement accuracy.  Conventionally, implementing the computing logical processes requires the engineering of dedicated interactions according to the characteristics of the encoding basis, which may require a specific physical setup, i.e. hardware, that cannot be changed as easily as the choice of encoding..  
The variety of encoding diversifies the architecture of CV quantum computers, and precludes the strengths of each encoding to be shared with all others.  

In this work, we describe two unified schemes for universal quantum computing with any CV encoding, in the sense that all logical processes are independent of the encoding state in each qumode.  Specifically, a qubit is stored in the parity of two or four qumodes.  All logical processes, which include computational state initialisation, universal set of logic gates, and state-readout, can be implemented by the preparation of encoding basis states, exponential-swap operations, and swap-tests.  We show how the required operations can be implemented with realistic physical interactions in CV systems.  Our schemes inherently allow logical qubits with different encoding to be brought to interact without decoding, thus the strength of different encodings can be utilised in the same computation.  Additionally, the logical states lie in the CV decoherence-free-subsystem, so the quantum information is robust against spatially collective noise.

\textit{Dual-Rail Scheme---} We begin by describing the simpler scheme in which each logical qubit consists of two qumodes.  For any encoding with the orthogonal single-mode basis states $|0_L\rangle$ and $|1_L\rangle$, where $\langle 0_L | 1_L\rangle=0$, we define the dual-rail logical ($D$-logical) basis states as $|0_D\rangle \equiv |0_L 1_L\rangle$ and $|1_D\rangle \equiv |1_L 0_L\rangle$.  A computation involving $N$ logical qubits thus requires $2N$ qumodes.  We denote that the $n$-th $D$-logical qubit is composed of the $(2n-1)$-th and the $2n$-th qumodes.  We assume the ability to efficiently prepare $|0_L\rangle$ and $|1_L\rangle$, and hence the initial computational state $|0_D\rangle^{\otimes N}$.

In the specific encoding of lowest energy Fock states, the $D$-logical states resemble the dual-rail single photon qubit in optical quantum computers \cite{Chuang:1995ud, Knill:2001vi}.  Nevertheless, the logic gate implementations of those schemes rely heavily on the property of the Fock basis states.  In stark contrast, the logic gates illustrated in this work are constructed from swap-based operations that are independent of the encoding bases.
By definition, the swap between the qumodes $i$ and $j$, $\hat{S}_{ij}$, is defined by the action $\hat{S}_{ij} \hat{a}_i \hat{S}^\dag_{ij} = \hat{a}_j$ for any $i$ and $j$, where $\hat{a}_i$ is the annihilation operator of qumode $i$.  The $D$-logical basis states are flipped by swapping the qumodes, i.e., $\hat{S}_{2n-1,2n}|0_D\rangle_n = |1_D\rangle_n$, so a swap plays the role of the Pauli-$X$ operation in the $D$-logical basis, i.e., $\hat{S}_{2n-1,2n} = \hat{X}_{D_n}$.

Arbitrary rotation on the $Y$-$Z$ plane of the $D$-logical Bloch sphere is accomplished by a two-mode exponential-swap (E-swap) operation \cite{Lloyd:2014gc,Lau:2016QML}, i.e., 
\begin{equation}
\exp(i \theta \hat{S}_{2n-1,2n}) = \exp(i \theta \hat{X}_{D_n})~.
\end{equation}
An entangling gate between two $D$-logical qubits could be a four-mode E-swap,
\begin{equation}\label{eq:DCX}
\exp(i\theta \hat{S}_{2n-1,2n} \hat{S}_{2m-1,2m}) = \exp(i \theta \hat{X}_{D_n}\hat{X}_{D_m})~.
\end{equation}

The universal set of logic gates can be completed by a phase-shift gate, $\exp(i \phi \hat{Z}_D)$ \cite{book:NielsenChuang}.  Unfortunately, we find that such an operation cannot be coherently implemented by any combination of swap.  Nevertheless, if a small gate error is permitted, the phase-shift gate can be efficiently implemented by the quantum machine-learning techniques \cite{Lloyd:2014gc}.  Consider that an ancillary qumode $c$ is prepared in $|0_L\rangle_c$.  After an E-swap operation is applied with the second qumode of the target $D$-logical qubit, the ancilla is then disposed.  For a small parameter $\epsilon$, the operation can be approximated as
\begin{eqnarray}\label{eq:DZ}
&&\textrm{Tr}_c\big\{e^{-i \epsilon \phi \hat{S}_{c, 2n}} \big(|0_L\rangle \langle 0_L|_c \otimes \rho_{D_n}\big) e^{i \epsilon \phi \hat{S}_{c, 2n}} \big\} \nonumber \\
&\approx& \big(e^{-i \epsilon \phi |0_L\rangle \langle 0_L|}\big)_{2n}\rho_{D_n}  \big(e^{i \epsilon \phi |0_L\rangle \langle 0_L|}\big)_{2n} + O(\epsilon^2)~.
\end{eqnarray}
The operation induces a phase shift $\epsilon \phi$ only on the $D$-logical qubit state $|1_D\rangle$.  Repeating the process for $1/(2\epsilon)$ times, we obtain the gate operation $\big(e^{-i 2 \phi |0_L\rangle \langle 0_L|}\big)_{2n}= e^{-i\phi}\exp(i\phi \hat{Z}_{D_n})$, with an overall error that scales as $O(\epsilon)$.  

After computation, the $D$-logical qubit is readout by a swap-test \cite{Filip:2002dq}.  Specifically, a controlled-swap operation, $\hat{\mathcal{C}}_{2n-1,2n} \equiv |0\rangle \langle 0|_A \otimes \mathbb{I}_{2n-1,2n} +|1\rangle \langle 1|_A \otimes \hat{S}_{2n-1,2n}$, where $\mathbb{I}$ is the identity operator, is applied to the $n$-th $D$-logical qubit and an auxiliary qubit $A$ that is prepared in $|+\rangle_A \equiv (|0\rangle_A + |1\rangle_A)/\sqrt{2}$. The total state becomes
\begin{eqnarray}
\hat{\mathcal{C}}_{2n-1,2n}|+\rangle_A |\psi_D\rangle_n &=&  |+\rangle_A (\mathbb{I}_{D_n}+\hat{X}_{D_n}) |\psi_{D}\rangle_n/2 \nonumber \\
 &&+|-\rangle_A (\mathbb{I}_{D_n}-\hat{X}_{D_n}) |\psi_D\rangle_n/2~,~
\end{eqnarray}
for any $D$-logical state $|\psi_D\rangle$.  Measuring the auxiliary qubit in the $X_A$ basis is effectively a projective Pauli-$X$ measurement on the $D$-logical qubit.  


\textit{Quad-Rail Scheme---} The main drawback of the dual-rail scheme is the incoherent implementation of the phase-shift gate that requires a surplus of operations and resources.  Here we describe a quad-rail scheme in which even the phase-shift gate can be implemented coherently.  We define the quad-rail logical ($Q$-logical) basis states as $|0_Q\rangle\equiv |+_D -_D\rangle$ and $|1_Q\rangle\equiv |-_D +_D\rangle$, where each consists of two $D$-logical qubits in either of $|\pm_D\rangle \equiv (|0_L1_L\rangle \pm |1_L0_L\rangle)/\sqrt{2}$.  We note that the scheme can be viewed as the concatenation of a variant of the exchange-interaction code on CV encodings \cite{Bacon:2000fy,Mohseni:2005di}.

A computation with $N$ $Q$-logical qubits requires $4N$ qumodes.  To initialise the computational state, 
the qumodes are first prepared in either of the encoding basis states, and then grouped into $2N$ pairs that each consists of one $|0_L\rangle$ and one $|1_L\rangle$.
A swap test is then applied to project each pair of qumodes to either $|+_D\rangle$ or $|-_D\rangle$.  
When $N$ is large, the probability of obtaining sufficient $|+_D\rangle$ and $|-_D\rangle$ for initialising $(1-\delta)N$ qubits of $|0_Q\rangle$ is exponentially close to unity in terms of $N$.

The $n$-th $Q$-logical qubit is composed of the $(4n-3)$-th to $4n$-th qumodes.  Swapping the first two qumodes of a $Q$-logical qubit will induce a phase of $-1$ only if the state is $|1_Q\rangle$; this operation plays the role of the $Q$-logical Pauli-$Z$ operator, i.e., $ \hat{S}_{4n-3,4n-2} = \hat{Z}_{Q_n}$.  Hence the single qubit phase-shift gate and the entangling conditional-phase gate can be implemented respectively by the two-mode and four-mode E-swap operations, i.e.,
\begin{eqnarray}\label{eq:QZ}
\exp(i\phi \hat{S}_{4n-3,4n-2}) &=& \exp(i\phi \hat{Z}_{Q_n})~, \\
\exp(i\phi \hat{S}_{4n-3,4n-2}\hat{S}_{4m-3,4m-2}) &=& \exp(i\phi \hat{Z}_{Q_n}\hat{Z}_{Q_m})~. ~\label{eq:QCZ}
\end{eqnarray}

The $Q$-logical basis states are flipped by swapping the first two qumodes with the last two qumodes of a $Q$-logical qubit.  This operation plays the role of the $Q$-logical Pauli-$X$ operator, i.e., $ \hat{S}_{4n-3,4n-1}\hat{S}_{4n-2,4n} = \hat{X}_{Q_n}$.  Hence the rotation gate on the $Q$-logical $Y$-$Z$ plane can be implemented by a four-mode E-swap operation,
\begin{equation}\label{eq:QX}
\exp(i\theta \hat{S}_{4n-3,4n-1}\hat{S}_{4n-2,4n}) = \exp(i\theta \hat{X}_{Q_n})~.
\end{equation}
Eqs.~(\ref{eq:QZ}), (\ref{eq:QCZ}), and (\ref{eq:QX}) complete the coherent implementation of the universal logic gate set in the $Q$-logical basis.

Similar to the dual-rail scheme, a $Q$-logical qubit is also readout by a swap-test.  An auxiliary qubit is first prepared in $|+\rangle_A$, and a controlled-swap operation is then applied to the first two qumodes of a $Q$-logical qubit, i.e., $\hat{\mathcal{C}}_{4n-3,4n-2}$.  The total state becomes
\begin{eqnarray}
\hat{\mathcal{C}}_{4n-3,4n-2}|+\rangle_A |\psi_Q\rangle_n &=&  |+\rangle_A (\mathbb{I}_{Q_n}+\hat{Z}_{Q_n}) |\psi_Q\rangle_n/2 \nonumber \\
 &&+|-\rangle_A (\mathbb{I}_{Q_n}-\hat{Z}_{Q_n}) |\psi_Q\rangle_n/2~.~~~
\end{eqnarray}
Measuring the auxiliary qubit in the $X_A$ basis is effectively a projective Pauli-$Z$ measurement on the $Q$-logical qubit.

Before proceeding, we emphasise that the logical processes of both the dual-rail and quad-rail schemes are implemented by the same set of operations: basis state preparations, two-mode and four-mode E-swap operations, and swap-tests.  As a comparison, implementing a $D$-logical phase-shift gate requires excessive operations and ancillae, but each $D$-logical qubit is composed of fewer qumodes, so the dual-rail scheme could be advantageous for a class of encodings that the phase-shift operation is efficient.  On the other hand, although each $Q$-logical qubit involves more qumodes, all logic gates can be implemented coherently for the $Q$-logical qubits.

\textit{Computation with Different Encodings---}  All the logical processes in the two schemes are constructed without specifying the encoding in each qumode.
Therefore, the sequence of physical operations that synthesises a computational circuit is the same, irrespective of the encoding basis states $|0_L\rangle$ and $|1_L\rangle$.  In fact, the computation can be conducted by a party that has no knowledge about the basis states except their orthogonality.

Examples of popular encoding basis states are shown in Table \ref{table:logical}.  The encodings are developed for their own purposes, but each of them inevitably suffers from certain drawbacks.  For instances, Fock state encoding \cite{Chuang:1995ud, Chuang:1997ko, Knill:2001vi} and coherent state encoding \cite{Ralph:2003jk,Anonymous:hr} enable efficient state preparation and linear-optical logic gates, but some logic gates are probabilistic and their implementations require stringent detection efficiencies.  Cat state encoding enables quantum error correction against photon loss \cite{Leghtas:2013ff,Mirrahimi:2014js}, but implementing the logic gates may require slow Zeno dynamics.  The Gottesman-Kitaev-Preskill (GKP) protocol enables fault-tolerant quantum computating and logical states to be readout by accurate homodyne detection, but the basis states are superpositions of squeezed states which the construction is technically challenging \cite{Gottesman:2001vk,Menicucci:2014cx}.

\begin{table}\caption{\label{table:logical} Examples of single-mode basis states of different encodings. The lowest energy Fock state is defined as $\hat{a}|0\rangle=0$. A coherent state with an amplitude $\alpha$ is given by $|\alpha\rangle = \hat{D}(\alpha)|0\rangle$, where $\hat{D}(\alpha)$ is the displacement operator \cite{VanLoock:2011kt}.  $|x\rangle_q$ is an infinitely squeezed state as well as an eigenstate of the $q$-quadrature, i.e., $\hat{q}|x\rangle_q = x|x\rangle_q$.  We have neglected the normalisation in the states.}
\begin{center}
\begin{tabular}{ |c|| c |c| }
\hline  Encoding & $|0_L\rangle$ & $|1_L\rangle$  \\ \hline
Fock state \cite{Chuang:1995ud, Chuang:1997ko, Knill:2001vi}\cite{note:Fock} & $|0\rangle$ & $\hat{a}^\dag|0\rangle$ \\ 
Coherent state \cite{Ralph:2003jk,Anonymous:hr} & $|\alpha\rangle$ & $|-\alpha\rangle$ \\ 
Cat state \cite{Leghtas:2013ff,Mirrahimi:2014js}& $|\alpha\rangle + |-\alpha\rangle$ & $|i\alpha\rangle + |-i\alpha\rangle$ \\ 
GKP \cite{Gottesman:2001vk,Menicucci:2014cx} & $\sum_{n=-\infty}^\infty |2n\rangle_q$ & $\sum_{n=-\infty}^\infty |2n+1\rangle_q$ \\ \hline
\end{tabular}
\end{center}
\end{table}

Remarkably, our schemes inherently allow logical qubits with different encodings to be brought to interact in the same computation, without the necessity to decode the quantum information.  This is because the four-mode E-swap operation, which implements the logical entangling gates in Eqs.~(\ref{eq:DCX}) and (\ref{eq:QCZ}), preserves the computational subspace of each $D$-logical qubits, i.e.,
\begin{equation}\label{eq:different_encoding}
\big[\exp(i\theta \hat{S}_{2n,2n-1}\hat{S}_{2m,2m-1}), \mathcal{I}_{D_n}(L_1)\otimes \mathcal{I}_{D_m}(L_2)\big]=0~,
\end{equation}
where $\mathcal{I}_D(L_i) \equiv |0_{L_i}1_{L_i}\rangle \langle 0_{L_i}1_{L_i}| + |1_{L_i} 0_{L_i}\rangle \langle 1_{L_i} 0_{L_i}| $ is the projection operator to the computational subspace of a $D$-logical qubit with the encoding $L_i$.  Because each $Q$-logical qubit is composed of two $D$-logical qubits, the property in Eq.~(\ref{eq:different_encoding}) also preserves the computational subspace of $Q$-logical qubits. 

In the dual-rail scheme, the same encoding is required only for the two qumodes in each $D$-logical qubit and the $O(1/\epsilon)$ ancillae required for implementing the phase-shift gate.  In the quad-rail scheme, it is sufficient for the four qumodes in each $Q$-logical qubit to have the the same encoding.

The logical processes of both the dual- and quad-rail schemes are sufficiently but not necessarily implemented by the swap-based operations; specific properties of an encoding could be used to achieve more robust or faster logical processes.  For example, a $D$-logical qubit with GKP encoding can be readout by homodyne detection, which could be more accurate than the qubit measurement in a swap-test.  Thus the strengths of different encodings can be utilised if each encoding is employed in the part of computation for which it is best adopted.   For instance, coherent states are efficiently created as undetected logical ancillae, cat states are best for transmitting quantum information through lossy links, and the final result is accurately readout from GKP qubits.

\textit{Decoherence Free Subsystem---}  
In CV quantum computation, leakage error is a major form of error as the environmental noise typically projects the encoded state out of the computational subspace.  In some CV systems, the noise is the same in each qumode.  Examples include the background magnetic field experienced by spin-ensembles, and the fluctuation of the trapping potential of an ion chain.  The decoherence effect of such collective noise can be reduced by storing the quantum information in the decoherence-free-subsystem (DFS) (also referred to as noiseless-subsystems \cite{note:DFS}) \cite{Knill:2000wt, Zanardi:2001vo, Kempe:2001bg}.
As a merit of our schemes, both the $D$-logical and $Q$-logical states are inherently within the DFS.  To the best of our knowledge, our schemes are also the first explicit protocols that incorporate DFS in CV systems.

The key idea is that collective noise commutes with the swap operation, i.e., $[\hat{U}_i \otimes \hat{U}_j, \hat{S}_{ij}]=0$, where $\hat{U}_i$ and $\hat{U}_j$ are the same unitary noise acting on different qumodes, so it also commutes with the E-swap and controlled-swap operations that constitute the logical processes.  For a computation that is implemented by a physical operation $\mathcal{U}$, the physical state that has suffered from subsequent collective noise is given by
\begin{equation}\label{eq:DFS}
\hat{U}^{\otimes M} \mathcal{U} |\Psi_0\rangle = \mathcal{U} \hat{U}^{\otimes M}  |\Psi_0\rangle~,
\end{equation}
where $|\Psi_0\rangle$ is the $M$-mode total initial state of the computational and the ancillary qumodes.  The right hand side of Eq.~(\ref{eq:DFS}) can be viewed as the implementation of the same computation
with the redefined encoding states, $|0_{L'}\rangle = \hat{U}|0_L\rangle$ and $|1_{L'}\rangle = \hat{U}|1_L\rangle$.  Because both $\mathcal{U}$ and the swap-test-based logical readout are the same for all encodings, the collective noise does not decohere the quantum information nor affects the computation result.

The DFS does not only protect the quantum information in the storage, it could also offer the protection during the logic gate operations.
If the controlled-swap, two-mode, and four-mode E-swap can be respectively implemented by the Hamiltonians $\hat{H}_\mathcal{C}\propto\hat{\mathcal{C}}$, $\hat{H}_2 \propto \hat{S}$, and $\hat{H}_4 \propto \hat{S}\otimes \hat{S}$, then the collective noise commutes with the logical processes, i.e., $[\hat{U}^{\otimes M},\hat{H}_\mathcal{C}]=[\hat{U}^{\otimes M},\hat{H}_2]=[\hat{U}^{\otimes M},\hat{H}_4]=0 $, and so the quantum information always lies within the DFS.  

Finally, we discuss about the order of correlation that is tolerable by the DFS.  
In the quad-rail scheme, because the same encoding is required only in each $Q$-logical qubit, $\hat{U}$ has to be identical among the four qumodes in the same $Q$-logical qubit.  On the other hand, although each $D$-logical qubit consists of only two qumodes, the same noise has to be experienced also by the $O(1/\epsilon)$ ancillae for implementing accurate phase-shift gates.  

\textit{Quantum Error Correction---} Some encodings enable quantum error correction (QEC) in each qumode \cite{Gottesman:2001vk,Leghtas:2013ff,Mirrahimi:2014js}.  
For an error channel $\mathcal{E}$ with an error $\delta$, a QEC process $\mathcal{R}$ can recover any encoding state $|\psi\rangle$, up to an error that scales as a higher order of $\delta$, i.e., $\mathcal{R}\big\{\mathcal{E}\{|\psi\rangle\langle \psi| \} \big\} \approx |\psi\rangle\langle \psi| + O(\delta^2)$.
Such a QEC process is applicable to correct errors in the $D$-logical qubit, which is essentially a two-mode encoding state, $|\psi\rangle = c_{00}|0_L0_L\rangle + c_{01}|0_L1_L\rangle+c_{10}|1_L0_L\rangle+c_{11}|1_L1_L\rangle$, with the constraint $c_{00}=c_{11}=0$.  Similarly, QEC is also applicable in the quad-rail scheme, because each $Q$-logical qubit stores quantum information in a subspace of the four-mode encoding states.

\textit{Physical Implementation---} The most ideal implementation of the E-swap and the controlled-swap operations is to apply the exchange-interactions in the form of $\hat{H}_2 \propto \hat{S}$, $\hat{H}_4 \propto \hat{S}\otimes \hat{S}$, and $\hat{H}_\mathcal{C}\propto \hat{\mathcal{C}}$, so that the computation is always conducted within the DFS.  To the best of our knowledge, however, such interactions have not been engineered in any CV system yet.  Alternatively, we propose an implementation that employs only realistic interactions, at the expense that the quantum information may leave the DFS during the operations.  

\begin{figure}
\begin{center}
\includegraphics{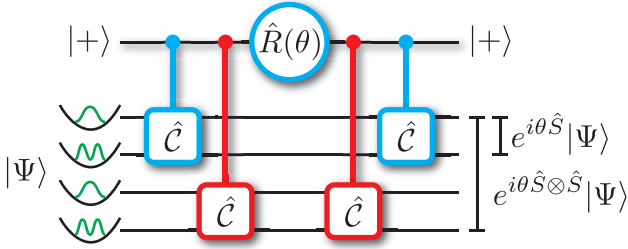}
\caption{ \label{fig:eswap} (Colour online) Circuit diagram of Eq.~(\ref{eq:eswap}) for implementing two-mode (blue only) and four-mode (blue + red) exponential-swap.  The ancilla qubit remains in $|+\rangle$ after the operations.}
\end{center}
\end{figure}

We require an auxiliary qubit $A$ that can be rotated on the $Y$-$Z$ plane of the Bloch sphere, i.e., $\hat{R}_A(\theta)=\exp(i\theta \hat{X}_A)$, to act as the control in the controlled-swap operation.   Initially, the qubit is prepared as $|+\rangle_A$.  The E-swap operations can be coherently implemented as \cite{Lau:2016QML},
\begin{eqnarray}\label{eq:eswap}
\hat{\mathcal{C}}_{ij}\hat{R}_A(\theta)\hat{\mathcal{C}}_{ij}|+\rangle_A |\Psi\rangle_{ij} &=& |+\rangle_A e^{i\theta \hat{S}_{ij}} |\Psi\rangle_{ij}~, \\
\hat{\mathcal{C}}_{ij}\hat{\mathcal{C}}_{kl}\hat{R}_A(\theta)\hat{\mathcal{C}}_{ij}\hat{\mathcal{C}}_{kl}|+\rangle_A |\Psi\rangle_{ijkl} &=& |+\rangle_A e^{i\theta \hat{S}_{ij}\hat{S}_{kl}} |\Psi\rangle_{ijkl}~, \nonumber \\
\end{eqnarray}
where $|\Psi\rangle_{ij}$ ($|\Psi\rangle_{ijkl}$) is the state of the qumodes $i,j$ ($i,j,k,l$).  The circuit diagram of the above procedure is shown in Fig.~\ref{fig:eswap}.

The implementation of the controlled-swap operation is best understood by expressing it as a unitary, 
\begin{equation}
\hat{\mathcal{C}}_{ij}= e^{-\frac{\pi}{4} (\hat{a}_i \hat{a}_{j}^\dag-\hat{a}_i^\dag \hat{a}_{j})} e^{i \frac{\pi}{2} (\mathbb{I}_A -\hat{Z}_A) \hat{a}^\dag_i \hat{a}_i} 
e^{\frac{\pi}{4} (\hat{a}_i \hat{a}_{j}^\dag-\hat{a}_i^\dag \hat{a}_{j})}~.
\end{equation}
This operation can be implemented by applying in a correct order the controlled-phase-shift interaction $\hat{H}_\textrm{K} \propto \hat{Z}_A \hat{a}^\dag \hat{a}$, and passive linear mode transformations that include beam-splitting, $\hat{H}_\text{BS}\propto \hat{a}_i \hat{a}^\dag_j + \hat{a}^\dag_i \hat{a}_j$, and phase-shifting, $\hat{H}_\textrm{P}\propto \hat{a}^\dag\hat{a}$ \cite{Filip:2002dq, Jeong:2014bk}.  While the linear transformations are prevalent, the interaction in the form of $\hat{H}_K$ can also be found in many CV systems, such as the dispersive coupling between a transmon qubit and a cavity \cite{Nigg:2012jj}, and the second-order magnetic field gradient that couples a diamond nitrogen-vacancy centre with a mechanical oscillator \cite{Ma:2016Cool}.  

\textit{Conclusion and Discussions---}  In this work, we have described a unified formalism to conduct universal quantum computation independently of the specific CV encoding of the qubit.  By storing the quantum information in the parity of two or four qumodes, all logical processes can be implemented by basis state preparations, exponential-swap operations, and swap-tests.  Both the $D$- and $Q$-logical states are inherently the decoherence-free-subsystem of the qumodes, so the stored quantum information is decoupled from collective noise.  Our schemes allow logical qubits with different encodings to brought to interact without decoding.  This unprecedented flexibility would allow the strengths of different encodings to be utilised in the same computation, when each encoding is employed in the computational process that it is best adopted, e.g. storage, logical transformation, transmission, or readout.  

We have proposed an implementation of the required exponential-swap and controlled-swap operations that involves only realistic interactions.  Nevertheless, we encourage experimentalists to engineer in CV systems the exchange-interactions $\hat{H}_2 \propto \hat{S}$, $\hat{H}_4 \propto \hat{S}\otimes \hat{S}$, and $\hat{H}_\mathcal{C}\propto \hat{\mathcal{C}}$, for implementing the computing processes within the DFS, in addition to the potential applications in quantum learning machines \cite{Lau:2016QML, Lloyd:2014gc,Rebentrost:2014fi}.  We anticipate possibilities in concatenating existing nonlinear Hamiltonians \cite{Lloyd:1999vz} or applying non-Gaussian measurements \cite{Andersen:2015dp, book:FurusawaVanLoock}.  In any case, once the exponential-swap and the controlled-swap operations are realised, they are sufficient for conducting universal quantum computation with any encoding.  Therefore, the encoding can be designed with a greater flexibility.  Our scheme could hence facilitate the development of the hybrid continuous-variable quantum computers \cite{VanLoock:2011kt,Andersen:2015dp, book:FurusawaVanLoock}.

\textit{Acknowledgement---} H.-K. L. thanks Christian Weedbrook for useful comments.  This work is supported by the Croucher Foundation, an Alexander von Humboldt Professorship, and the EU projects EQUAM and QUCHIP.




\bibliographystyle{phaip}
\pagestyle{plain}
\bibliography{UUQC_bib}

\begin{thebibliography}{10}
\bibitem{note:Fock}
Here the two Fock basis states are given only as a possibility. Typical Fock
  state schemes employ dual-rail encoding that a photon is in either one of two
  qumodes. Nevertheless, our schemes can be trivially extended to the dual-rail
  Fock qubits by using four (eight) qumodes in each $D$-logical ($Q$-logical)
  qubit, and replacing the two-mode (four-mode) E-swap operations with the
  four-mode (eight-mode) E-swap.
\bibitem{note:DFS}
Here the decoherence-free-subsystems should not be confused with
  decoherence-free-subspaces. For the later, the encoded state remains
  unchanged by collective noise. In our case of decoherence-free-subsystem,
  however, the collective noise can change the state except the degree of
  freedom that stores the quantum information.
\end{thebibliography}

\end{document}